 \documentclass[final,3p,times,sort&compress]{elsarticle}

\usepackage[dvipsnames]{xcolor}
\usepackage[breaklinks,colorlinks,citecolor=blue,linkcolor=Green,urlcolor=blue]{hyperref}
\usepackage{amssymb}
\usepackage{gensymb}
\usepackage{float}

\journal{Nuclear Physics B}

\begin{document}

\begin{frontmatter}



\title{Independent check of sporadic beta decay anomalies reported earlier by Parkhomov}

\author[1]{Andrei E. Egorov\corref{cor}}
\ead{aegorov@runbox.com}
\author[2,3,4]{Aleksey A. Alekseev}
\ead{regulred25@gmail.com}
\cortext[cor]{Corresponding author}

\affiliation[1]{organization={Nuclear Physics and Astrophysics Division, Lebedev Physical Institute},
            addressline={Leninskii prospect – 53}, 
            city={Moscow},
            postcode={119333},
            country={Russia}}
\affiliation[2]{organization={Physics Department, Lomonosov Moscow State University},
	addressline={Leninskiye Gory – 1}, 
	city={Moscow},
	postcode={119991},
	country={Russia}}
\affiliation[3]{organization={Kulakov National Medical Research Center For Obstetrics, Gynecology And Perinatology},
		addressline={Oparina street -- 4}, 
		city={Moscow},
		postcode={117997},
		country={Russia}}
\affiliation[4]{organization={Russian Research Institute of Health},
	addressline={Dobrolyubova street -- 11}, 
	city={Moscow},
	postcode={127254},
	country={Russia}}

\begin{abstract}
A. Parkhomov reported in the past the detection of strong, short, sporadic, nearly everyday decay rate increases in $\beta$-isotopes placed at the focus of a concave mirror directed at a clear sky. He interpreted this effect as neutrino-induced decay. These neutrinos were assumed to be slow and arrive from space in the form of gravitationally collimated beams. We precisely replicated this experiment for the purpose of independent verification. We recorded about 50 days of data employing Geiger-counter-based signal and control detectors, $^{40}$K and $^{90}$Sr/$^{90}$Y isotopes as $\beta$-sources. We did not detect any significant difference between the data from signal and control detectors, both datasets obeyed the standard Poissonian process. Thus, we robustly excluded the hypothesized new effect. We suspect that A. Parkhomov was just misled by a trivial electromagnetic instability of his setup.  
\end{abstract}



\end{frontmatter}


\section{\label{sec:i}Introduction and motivation}

Dr. Alexander Parkhomov (AP) reported in the past the discovery of a new physical effect: $\beta$-decaying isotopes, which are placed at the focal point of a concave parabolic mirror pointed at the clear sky, manifest frequent and strong increases in their decay rates. This anomaly was described in \cite{2018JMPh....9.1617P,2011JMPh....2.1310P,Parkhomov-ru,Parkhomov-en,2010arXiv1010.1591P,Parkhomov-IJPAP}. When we refer to AP through our paper, we mean these publications, where AP presented in detail both his empirical findings and original theoretical explanation of the observed effect. However, no one has yet reported any attempts to replicate independently the anomalies claimed by AP. And the authors of this work conducted such replication, the results of which are reported here.

In this section we \textit{briefly} expound the whole AP's concept (to the best of our understanding) \textit{without} any substantial critical assessment. For more details we refer readers to the publications cited above. Then in section \ref{sec:our} we describe our precise replication of the original AP's experiment. Section \ref{sec:res} reports the results. Section \ref{sec:dis} discusses the experiment in a broad context. And the last section \ref{sec:con} summarizes our work.

A while ago, AP suspected that some yet-unknown cosmic agent manifests itself by influencing the decay rate of $\beta$-radioactive substances. AP conducted a series of experiments, which led him to the following conclusions about the hypothesized agent: this should be a neutrino-resembling elementary particle (let us denote it by $\nu_x$), which comes from space (sometimes in the form of collimated flows), interacts with matter through (at least) the weak force and possesses wave properties. The former property obviously follows from the fact of influence on $\beta$-radioactivity, and the latter one was deduced from observation of the phenomenon of $\nu_x$ diffraction on periodic structures. This observation motivated AP to make the next step in his experimentation -- he created a series of telescopes designed for $\nu_x$ detection. These telescopes utilized the assumed wave behavior of $\nu_x$: they contain a concave reflecting mirror, which is supposed to focus the plane-parallel $\nu_x$ wavefront analogously to how optical astronomical telescopes focus photons, and the detector in the form of $\beta$-decaying source paired with Geiger counter (GC). The reflection of $\nu_x$ from the mirror surface is assumed to happen through elastic scattering process. The detection of $\nu_x$ is assumed to happen through the capture process by $\beta$-decaying nuclei with subsequent emission of $e^\pm$, which are detected in turn by GC:
\begin{equation}\label{eq}
	\nu_x|\bar{\nu_x} + N(A,Z) \rightarrow N(A,Z \mp 1) + e^\pm. 
\end{equation}
Such inverse $\beta$-decay reaction does not have any $\nu_x$ energy threshold in the case of spontaneously decaying $N(A,Z)$. Produced $e^\pm$ have the fixed distinct kinetic energy: it is easy to show that their kinetic energy exceeds the $\beta$-decay spectrum endpoint by $2m_{\nu_x}c^2$ (e.g., \cite{2023JCAP...06..021C}). Thus, according to AP, $\beta$-decaying isotopes, which were placed at the mirror focus point, manifested frequent and strong activity increases, which were distributed quasi-randomly in time and mirror direction on the sky. AP reported positive results with $^{40}$K, $^{60}$Co, $^{90}$Sr/$^{90}$Y and even with a bare GC. The latter case was interpreted as decay rate bursts of trace $\beta$-active nuclides in the body of GC and ambient air (e.g., radon). One may naturally suspect the presence of trivial electrical noises in this situation instead of the actual activity increases. AP excluded such possibility by claiming two following facts. One is the classical controlling scheme, when the second identical detector works in parallel with the main one but is placed aside from the mirror. The control detector did not show any anomalies. The second reported fact is absence of the anomalies during cloudy weather. Local electrical interferences are not expected to have such anticorrelation. These facts convinced AP of the realness of anomaly, as well as led him to the conclusion, that atmospheric clouds attenuate cosmic $\nu_x$ flows.

Let us now write out AP's findings more quantitatively. His most efficient telescope has the steel mirror with parabolic surface ($\sim 10~\mu$m surface quality), diameter of 22 cm, focal length of 10 cm and thickness of $\sim$ 1 cm. The characteristic size of $\beta$-decaying targets is $\sim$ 0.1 cm. For electron detection, AP \textit{mainly} used the miniature industrial GC called "SBM-12" (GC-12 hereafter for simplicity) with $\sim$ 0.1 cm$^2$ sensitive zone covered by a metal with $\approx$ 0.005 cm thickness, but other detectors were occasionally employed too. The mirror with attached detector was situated on a simplistic equatorial mount with an electric motor, which can swing the mirror along the declination sky coordinate in the range $0\degree \lesssim \delta \lesssim 40\degree$ for efficient sky surveying. Thus, employing Earth's rotation, such telescope is able to sweep rather large portion of the sky during a day. After doing such sky scanning for several years, AP revealed the following key properties of the purported signals. The rate of statistically significant signals was $\sim$ 1 day$^{-1}$ (meaning clear days). The typical signal duration is $\sim$ 1 s, though signals lasting minutes to hours and having fine temporal structure were detected too. The signal amplitude was up to $\sim$ 1000 times (!) with respect to the steady count rate; which was $\sim$ 1 s$^{-1}$, and was composed of a minor contribution from external background and the main contribution from the normal decay rate of the $\beta$-source.

After decades of his research, AP arrived to the following theoretical explanation of the observed effect: it is caused by the electron neutrinos with the mass $m_{\nu_e} \sim$ 10 eV, which float in the space like the cosmic neutrino background (C$\nu$B) with typical speeds of hundreds km/s. Such large value of $m_{\nu_e}$ was considered; since its limits from direct measurements back to the times, when AP was conducting his research in 1990--2010, were not so stringent \cite{2001NuPhS..91..280L} as nowadays. The possibility of $\nu_e$ significant interactions with matter was justified by AP through the fact of very large de-Brogile wavelength $\lambda_{dB} \equiv 2\pi \hbar /p$ of such particles -- $\lambda_{dB}$ values fall into $\mu$m--mm range for $\nu_e$ parameter values mentioned above. He assumed that the typical weak interaction cross section is enhanced in this situation proportionally to the number of target nucleons in the volume $\sim \lambda_{dB}^3$, which is rather macroscopic. Moreover, AP drew the analogy between $\nu_e$ and photons with wavelengths $\lambda_{dB}$, and treated $\nu_e$ by laws of wave optics. Such treatment was initially motivated by the mentioned diffraction observation and then justified $\nu_e$ focusing by the concave mirrors. Another essential element of AP's concept is dense and narrowly collimated $\nu_e$ beams, which are absent "by default" in C$\nu$B. AP invented such beams through the mechanism of local gravitational focusing of globally uniform and isotropic $\nu_e$ background by the Sun and nearby stars. An example of modeling of such focusing can be seen in e.g. \cite{2014ApJ...780..158P}. Essentially, this mechanism creates dense and anisotropic structures like streams and caustics (meanwhile, similar mechanism was also considered in \cite{2017PDU....17...13B}). AP suggested that an ambient uniform $\nu_e$ background is not dense enough in order to cause a detectable reaction of his telescope. But sometimes the Sun and nearby stars create short-lived, non-stationary, narrowly-directed $\nu_e$ flows with the density by many orders of magnitude larger than that of background. Such flows strike the telescope, get focused by its mirror and cause $\beta$-radioactivity bursts at the detector. According to AP, the gravitational focusing works especially efficiently; when the Earth, Sun and some nearby star get aligned along the straight line. At such moments we see the Sun conjunction with that star in our sky. AP reported certain reproducibility of the signals recorded at the days of some of such conjunctions in different years. Finally, based on estimates of $\nu_e$ spatial concentration, AP suggested that these particles can account for all dark matter (DM) (at least in our Galaxy) rather than just C$\nu$B.

AP's concept, which we expounded above, is indeed crude and vague in a multitude of aspects. First of all, the modern constraints on $m_{\nu_e}$ from both direct measurements and cosmological observations yield $m_{\nu_e} \lesssim 0.1$ eV \cite{2024arXiv240613516A}, thus excluding the value estimated by AP. Eliminating the usual, i.e. active, electron neutrinos from consideration, one may in principle hypothesize a manifestation of sterile neutrinos \cite{2023JCAP...06..021C} here; which are assumed to be heavier than $\nu_e$, may have a small mixing with $\nu_e$ and represent the viable DM candidate. Having such perspective in mind, the authors of the present work got inspired to \textit{check only the empirical part} of AP's work \textit{fully agnostically} of theoretical possibilities. Saying by another way, we aimed to check independently whether the claimed $\beta$-radioactivity anomaly is reproducible and real.

\section{\label{sec:our}Our replication of Parkhomov's experiment}

\begin{figure}[h]
	\centering
	\includegraphics[width=0.5\linewidth]{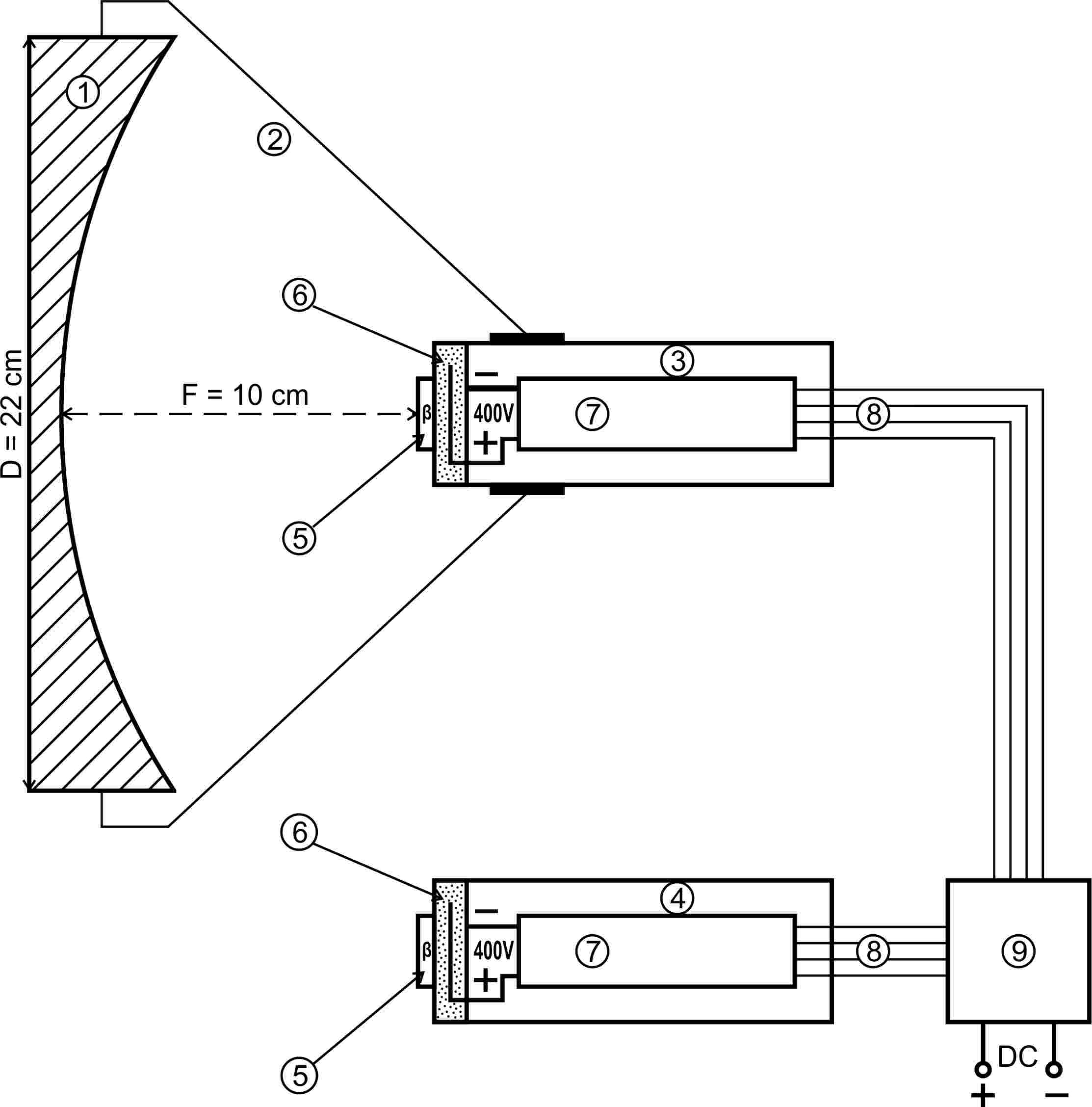}
	\caption{\label{fig:scheme}The principal scheme (side view) of our experimental setup: 1 -- the steel mirror with parabolic surface; 2 -- four supports; 3 -- the main (signal) detector; 4 -- the control detector; 5 -- $\beta$-sources; 6 -- Geiger counters; 7 -- powering and linking electronic boards; 8 -- I2C bus; 9 -- the main electronic unit containing CPU, display, memory card etc. The relative sizes in the figure are close to reality.}
\end{figure}

Our practical goal was to replicate AP's setup as identically as possible. We were communicating directly with AP during our experimentation, and he kindly gave us his mirror with mount described in the previous section. We created new detectors telic for our purposes. We conducted our measurements during the summer seasons of 2023 and 2024. The first season was preparatory to some degree: we were testing various detector configurations, and understanding their noises and backgrounds. Thus, we drew our final conclusion based majorly on the measurements made in 2024, when our setup was firmly established. This setup is depicted in fig. \ref{fig:scheme}. Two nearly identical detectors were always working: the signal detector with $\beta$-decaying source fixed at the mirror focal spot and the control detector (with the same source) put aside from the mirror. For $\beta$-sources we employed two (enveloped) isotopes: $^{40}$K in the form of natural potassium chloride and $^{90}$Sr/$^{90}$Y with the activity $\approx$ 40 Bq (measured by GC). Both sources have the cylindrical (coin-like) shape with diameter $\approx$ 2 cm and height $\approx$ 0.2 cm. These coin-shaped sources were positioned orthogonally to the mirror's "optical" axis in order to efficiently intercept the hypothetical focused $\nu_x$ beam. The source thickness along the axis was chosen as a compromise between sufficient source activity and smallness of electron absorption (we aimed to detect only electrons from the reaction (\ref{eq}) as AP did it). These sources were fixed in front of the thin mica window of GC with diameter $\approx$ 4 cm. We employed mainly GCs called "Beta-1" \cite{beta-1}. Following AP's recipe, we inserted the additional $\approx$ 0.5 cm layer of plastic between the source and GC window in the case of $^{90}$Sr/$^{90}$Y in order to filter out the majority of "normal" electrons (including all the electrons from $^{90}$Sr) and thus increase sensitivity, since all the target electrons are expected to have the same (high) energy near the $\beta$-spectrum endpoint. $^{40}$K source is relatively weak, hence no additional attenuation was used. GCs were powered by 400 V potential with several layers of voltage stabilization. This setup yielded the following steady mean count rates: 2.3 s$^{-1}$ for $^{40}$K and 1.5 s$^{-1}$ for $^{90}$Sr/$^{90}$Y. These rates are dominated by counts from the sources, the external background rate is $\lesssim$ 0.5 s$^{-1}$.

The data from both detectors, i.e. the number of accumulated counts, were recorded on the memory card simultaneously at 1 and 10 Hz rates in order to have the opportunity to analyze data with various temporal resolutions. Two different sky survey methods were utilized. The ordinary method was rather trivial: the telescope mount was just fixed to the stationary position, and the motor was swinging the mirror back and forth with the speed $\approx 7\degree$/min and amplitude $\approx 40\degree$. The second survey mode was used during the days of Sun conjunction with nearby stars or clusters in the sky. As was mentioned in section \ref{sec:i}, AP stated that on such days, especially strong signals are expected. Quantitatively, the angular separation between the solar disk center and the star must not exceed several arcmin, and the gravitationally collimated $\nu_x$ beams are anticipated to arrive from the region with $\approx 15\degree$ radius around the Sun. We were scanning this region around the Sun semi-manually at such days. Specifically, we picked those conjunctions, when the stars with visual magnitudes brighter than 6$^m$ went behind the solar disk, i.e. the angular separation was $\leqslant 15'$.

Besides the main detector configurations described above, we also occasionally used others. Thus, in 2023, we recorded several days using miniature NaI-based dosimeters paired with $^{40}$K targets. Then in 2024 we recorded a small dataset employing bare GC-12 during the Sun conjunction with Messier 44 open star cluster. Table \ref{table} summarizes our accumulated recording times in various modes under reasonably clear sky. We can see that, including preliminary 2023 data, we recorded in total about 27 days with $^{40}$K and 24 days with $^{90}$Sr/$^{90}$Y. The conjunctions, which satisfy the criteria above, were happening during 16 days, which intersected in time with our recordings. As we mentioned in section \ref{sec:i}, AP observed significant signals nearly every clear day. Thus, our exposure in every mode is absolutely sufficient in order to draw a firm conclusion on the hypothesized effect. The former follows in the next sections.
\begin{table}[h]
	\caption{\label{table}An estimate of our recorded exposure, i.e. the number of clear days, when the telescope was working. Numbers in brackets mean the number of days, when the exposure intervals were intersecting in time with conjunctions of Sun with nearby stars. More details are in section \ref{sec:our}.}
	\centering
	\begin{tabular}{cccc} 
		\hline
		\noalign{\vskip 1pt}
		Year & $^{40}$K & $^{90}$Sr/$^{90}$Y & GC-12 \\ 
		\hline
		2023 & 2.0 with NaI, 13 with GC (3) & 0 & 0 \\ 
		2024 & 12 (3) & 24 (7) & 0.50 (3) \\ 
		\hline
	\end{tabular}
\end{table}

\section{\label{sec:res}The results of our measurements.}

Let us start here from the note on what kind of signals we were aiming to reveal. AP did not elaborate very well on the event occurrence rate, and their distribution by duration and intensity. But roughly, to the best of our understanding, we expected the following minimal occurrence rates in terms of orders of magnitude normalized to the steady $\beta$-source activity: one 10-fold event (i.e., increase or burst) every day, one 100-fold event every 10 days and one 1000-fold event every 100 days (meaning clear days). These signals are expected to have duration $\gtrsim$ 1 s and to be rather insensitive to the $\beta$-isotope. As we outlined in the previous section, our GC mean count rate, which reflects the steady/normal $\beta$-source activity, is $\sim$ 1 s$^{-1}$. Thus, taking into account our total exposure, we may anticipate tens of $\gtrsim$ 10 s$^{-1}$ events and even several $\sim$ 100 s$^{-1}$ events above the normal Poissonian background in our data even with the modest 1 Hz time resolution. 

\begin{figure}[t]
	\centering
	\includegraphics[width=0.497\linewidth]{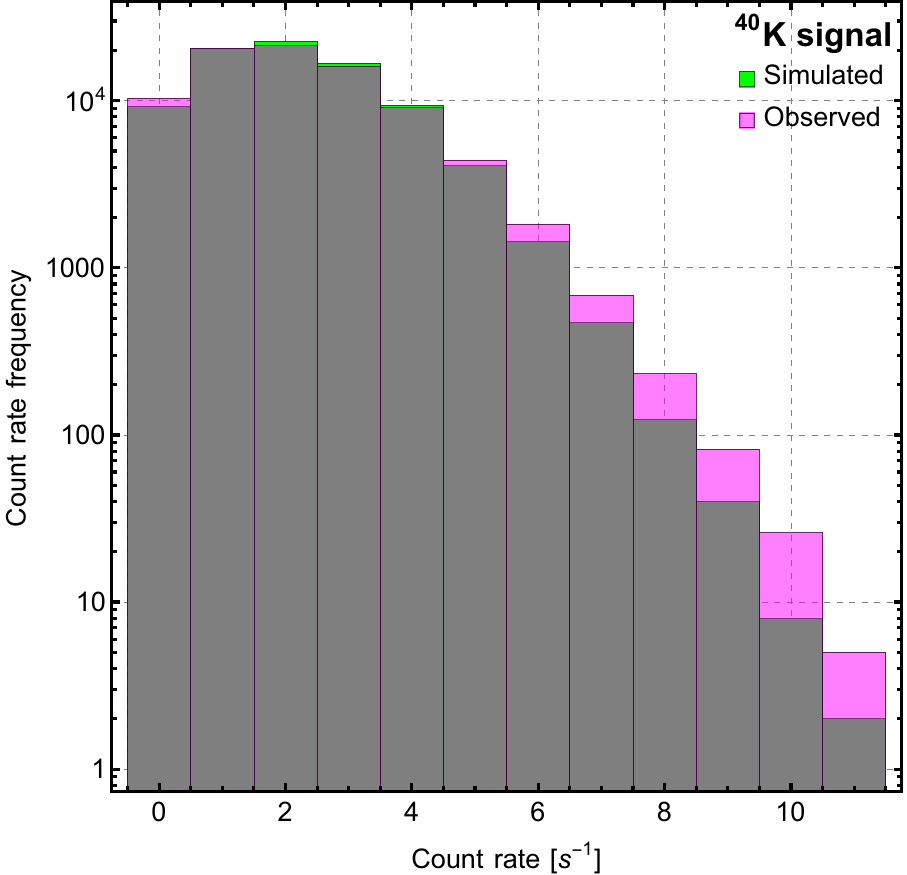}
	\includegraphics[width=0.497\linewidth]{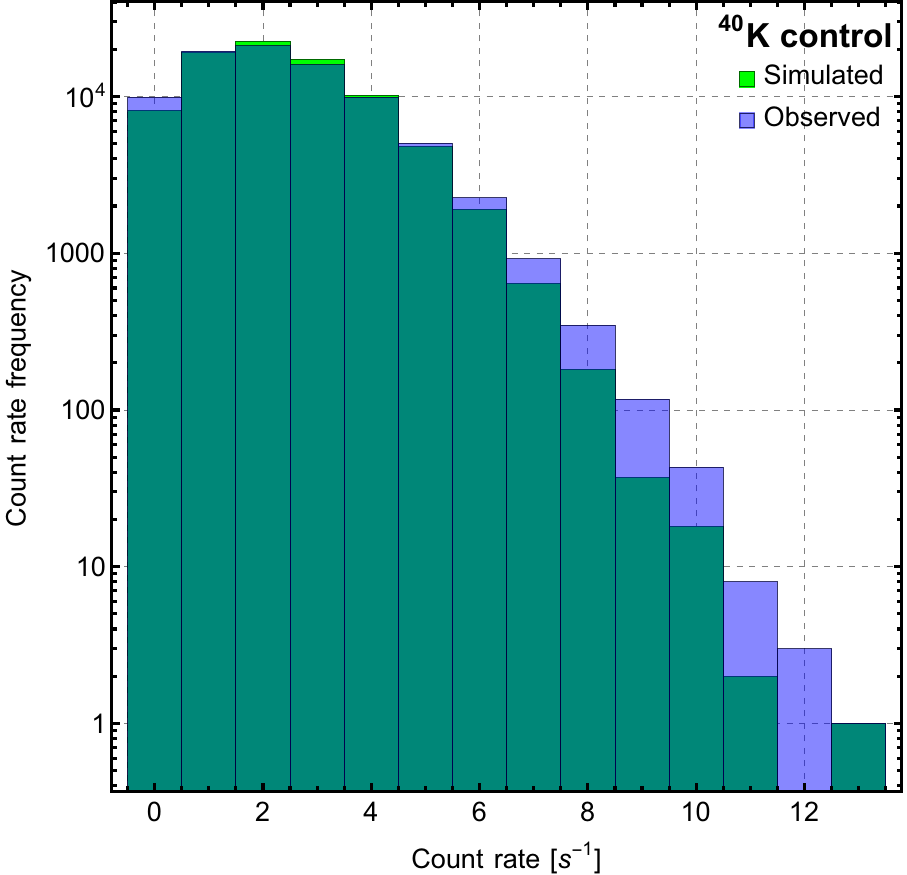}
	\caption{\label{fig:K}The measured and simulated count rate distributions (count integration time is 1 s) for the case of $^{40}$K source. The measured data was gathered during the whole clear day of 8th June 2024. The left panel reflects the signal detector at the mirror focus, the right panel reflects the control detector aside. The darker (bottom) bar sections represent the overlap of two datasets.}
\end{figure}

\begin{figure}[h]
	\centering
	\includegraphics[width=0.497\linewidth]{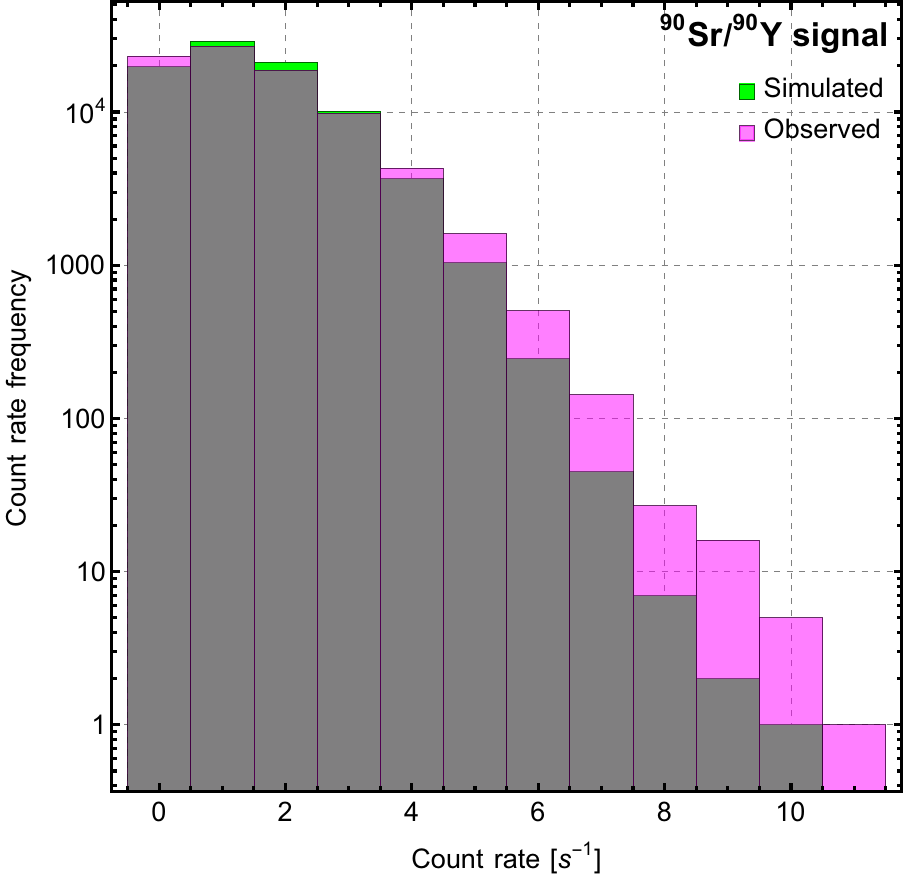}
	\includegraphics[width=0.497\linewidth]{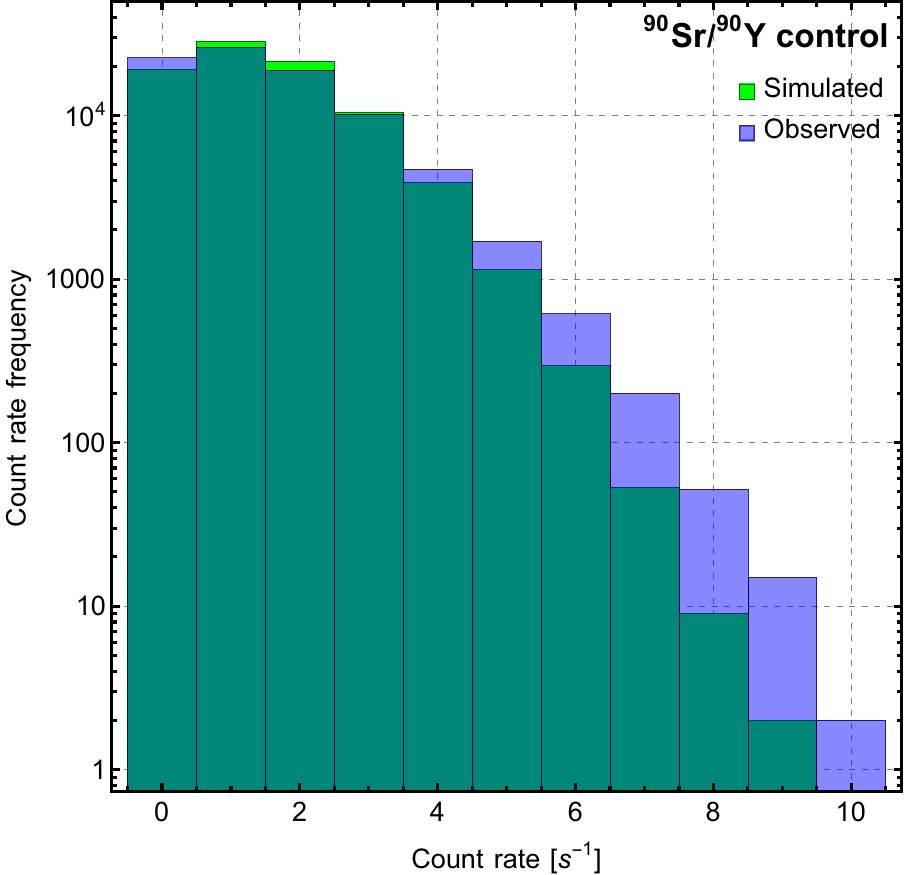}
	\caption{\label{fig:SrY}Same as fig. \ref{fig:K}, but for $^{90}$Sr/$^{90}$Y source measured on 28th June 2024.}
\end{figure}

Examples of our daily data for both $\beta$-isotopes are shown in fig. \ref{fig:K}-\ref{fig:SrY}. These histograms represent the numbers of occurrences of various observed counts integrated over 1 s during the whole day. We also plotted there the corresponding computer-simulated random realizations of Poisson distribution for comparison. These particular examples can be considered as representative, since the data from all other days do not differ significantly from these trial days. We can see at first a good accordance between the signal and control count rate distributions. At second, they both agree well with the simulated Poissonian count rate distributions. We can notice just a mild heavy tail in the observed distribution of both signal and control counts. We attribute such tail to an imperfectness of insulation of our detectors from nuisance electromagnetic interference, which may create a relatively small number of false counts.

Over all $\approx$ 50 days of our exposure, we saw something abnormal only once. This happened on 17th August 2023, when $^{40}$K count rate jumped up by $\sim$ 100 times for 1--2 s. However, we attributed this candidate event to the bogus signal rather than real, since at that time we had not yet achieved full understanding of possible electrical noises in our relatively simple setup. Once we eliminated all possible electromagnetic interference, no significant signals were observed above the steady rate. Also, if we assume that burst to be real, then it must not be so alone. As we outlined above, according to AP we should have tens of events with amplitudes above 10 s$^{-1}$. But nothing significant was seen beyond the steady Poissonian decay rate over all the robust data gathered in 2024 for both tested isotopes at the signal detector.

AP in fact was recording his data in a different way due to various reasons. He was recording not the number of counts during fixed time intervals, but instead the time interval durations with the fixed number of accumulated counts (usually $N$ = 256). Then he was deriving the count rate through division of $N$ by the accumulation duration. Such method yields a bit integrated signal and erases weak features. Thus, for example, AP was recording $^{40}$K with $N$ = 128 during 4 calendar months and detected 34 bursts with amplitudes of $\geqslant$ 2-fold with respect to the steady count rate. We processed our data through this method too for a comprehensive comparison. This algorithm yielded at most $\approx$ 1.5 times increases, which were the same on the signal and control detectors. Therefore, the absence of anomalies was confirmed. Indeed, the signal amplitudes are not equivalent in two registration methods.

\begin{figure}[h]
	\centering
	\includegraphics[width=0.497\linewidth]{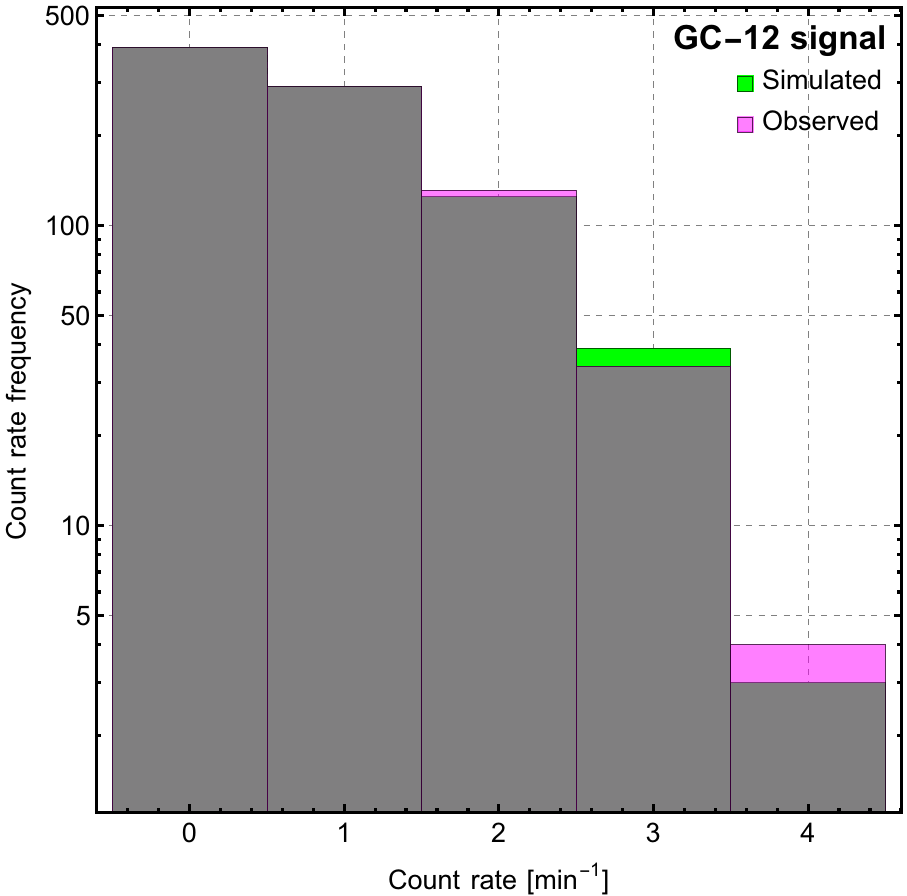}
	\includegraphics[width=0.497\linewidth]{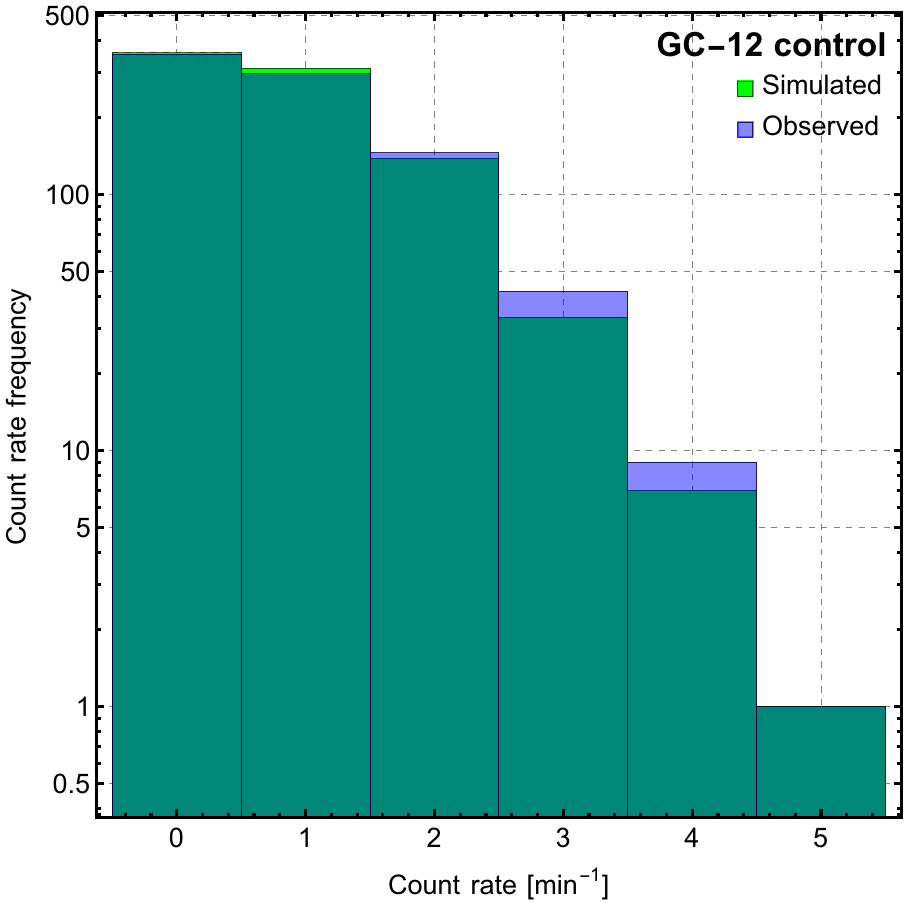}
	\caption{\label{fig:12}The measured and simulated count rate distributions (count integration time is 1 min) for the whole exposure ($\approx$ 12 hours) recorded with bare GC-12 on 29--31st July 2024 during the Sun passage near M44 star cluster on the sky.}
\end{figure}

Finally, we also tested AP's claim about strong signals obtained with bare GC-12 near the Sun, when it passes nearby M44 open star cluster in the sky. This was happening on 29--31st July 2024. The weather was not very good, but we were able to record $\approx$ 12 hours of data under sunshine, which should be sufficient to see at least something. But absolutely no hints of anomalies were found again. The corresponding count rate distributions are depicted in fig. \ref{fig:12}. We can see nearly perfect agreement between the measured and simulated distributions.

\section{\label{sec:dis}Discussion}

We suppose that we replicated AP's experimental setup identically enough and gathered sufficient amount of data in order to draw the reliable verdict for the effect being tested. We suspect that AP was observing just trivial electromagnetic interference/noise during his work. The presence of anomalies on his signal detector only and their absence on the control detector could be due to a physical/electrical contact between the signal detector mounted on the mirror and the whole telescope, which includes electrical motor etc. The claimed anticorrelation between the signals and atmospheric clouds might be a result of psychological bias. Thus, for example, in-depth analysis of AP's data revealed strong and unphysically looking signals during the night on 3-4 August 2007, when a strong rain with thunderstorm was going on in Moscow (where AP was working), according to the weather archive. This example supports the above reasoning. Additional suspicion arises from the absence of specific inherent characteristic timescale in hypothesized signals: they were reported to last from seconds to hours, which resembles just everyday human activity timescales.

Speaking more about experimental weaknesses, AP almost solely relied on just one detector type -- i.e. GC, which is known to be a primitive and noise-vulnerable particle detector \cite{2022SSRv..218...64P}. Another possible weakness is that AP powered his setup by a usual city electrical network and never tried autonomous power supply.

On the theoretical side, AP did not analyze enough unavoidable manifestations of the hypothesized particle and its effects beyond his narrow concept. Thus, the following simple thought experiment severely challenges the concept. AP estimated the coefficient of reflection of $\nu_x$ from the telescope mirror surface to be $k \sim 0.01$. In the absolute units, this corresponds to a tiny area of $\approx$ 4 cm$^2$. This essentially represents the effective area of his detector. This implies that any $\beta$-source alone with such size under the clear sky inevitably must demonstrate similar activity bursts with even larger occurrence rate, since the mirror collects $\nu_x$ flows from a small portion of sky. But such bursts do not exist in reality, obviously; and even if $k = 1$, this effect can not stay unnoticed nowadays. Thus, for example, the experiment \cite{2015APh....61...82B} watched $\approx$ 6.2 kg of the natural potassium during about 500 days above ground and found no decay rate anomalies down to the relative amplitude $\sim 10^{-5}$. In this respect, the mirror seems to play a crucial role only for the purpose of fine sky imaging, but not for just detection/revelation of hypothesized $\nu_x$. In order to overcome this challenge, one needs to make unusual assumptions: e.g., collective effects in $\nu_x$ beam, when its efficiency of $\beta$-decay induction depends on the flux non-linearly, and the beam compression by the mirror provides non-linear induction enhancement.

Besides sporadic bursts, AP also reported in \cite{2011JMPh....2.1310P,2018JMPh....9.1617P,Parkhomov-en,Parkhomov-ru,2010arXiv1010.1591P} the detection of small annual oscillation of $\beta$-decay rate with relative amplitude $\sim 0.1\%$ by watching various isotopes by various GCs for over 10 years. He attributed this modulation to $\beta$-decay induction modulation due to minor annual changes in the smooth background flux of hypothetical $\nu_x$. In our work, we did not aim to test this minor slow oscillation. However, the work \cite{2022NatSR..12.9535P} was dedicated to analysis of the former. These authors found out a strong anticorrelation between the ambient air humidity in Moscow shifted forward in time by $\sim$ one month and the decay rate in AP's data. Thus, they concluded that the claimed abnormal oscillation can be likely due to just instability of measurement conditions, which often causes such false detections \cite{2024Metro..61a5001P}. This conclusion can be ambiguous; because the air humidity outside, i.e. at some city weather monitor, can differ significantly from the air humidity in immediate vicinity of the experimental setup. But if the conclusion is valid, our work neatly complements \cite{2022NatSR..12.9535P}: we can exclude all the $\beta$-decay anomalies claimed by AP! In general, S. Pomme et al. conducted very extensive and valuable work on the critical assessment of the claims about the influence of cosmic neutrinos on $\beta$-decay \cite{2022SSRv..218...64P}.

\section{\label{sec:con}Conclusions}

We conducted independent and precise replication of AP's experiment, when $\beta$-decaying sources are being placed at the focus of steel concave mirror pointed at a clear sky. AP reported in the past an observation of sporadic and strong activity increases in this setup. We recorded about 27 days of data with $^{40}$K and 24 days with$^{90}$Sr/$^{90}$Y employing GCs "Beta-1". We did not observe any significant differences between the data from signal and control detectors. The count rate in both datasets follows well the standard Poisson distribution. Thus we robustly excluded the hypothesized effect and related necessity of new physics. Nevertheless, such brave AP's attempt to attack one of the biggest puzzles in modern science -- i.e., the physical nature of DM -- being worked alone with very scarce resources definitely deserves respect! 

Based on \cite{2022SSRv..218...64P,2024Metro..61a5001P}, we can probably state the global conclusion, that currently no sufficiently justified deviations from the standard exponential law of radioactive decay have been found.

\section*{Acknowledgements}

We acknowledge helpful usage of the following software in our work: R programming language, RStudio, Wolfram Mathematica, Stellarium and TeXstudio.


\bibliographystyle{elsarticle-num} 
\bibliography{~/YD/DM/universal}



\end{document}